\documentclass[english]{revtex4-1}
\usepackage[T1]{fontenc}
\usepackage[latin9]{inputenc}
\usepackage{float}
\usepackage{amsmath}
\usepackage{amssymb}
\usepackage{graphicx}
\usepackage{wasysym}
\usepackage{babel}
\begin{document}

\title{Prediction of deviations from the Rutherford formula for low-energy
Coulomb scattering of wavepackets}

\author{Scott E. Hoffmann}

\address{School of Mathematics and Physics~\\
The University of Queensland~\\
Brisbane, QLD, 4072~\\
Australia}
\email{scott.hoffmann@uqconnect.edu.au}

\begin{abstract}
We calculate the nonrelativistic scattering of a wavepacket from a
Coulomb potential and find deviations from the Rutherford formula
in all cases. These generally occur only at low scattering angles,
where they would be obscured by the part of the incident beam that
emerges essentially unscattered. For a model experiment, the scattering
of helium nuclei from a thin gold foil, we find the deviation region
is magnified for low incident energies (in the keV range), so that
a large shadow zone of low probability around the forward direction
is expected to be measurable.

From a theoretical perspective, the use of wavepackets makes partial
wave analysis applicable to this infinite-range potential. It allows
us to calculate the everywhere finite probability for a wavepacket
to wavepacket transition and to relate this to the differential cross
section. Time delays and advancements in the detection probabilities
can be calculated. We investigate the optical theorem as applied to
this special case.
\end{abstract}
\maketitle

\section{Introduction}

It is well known that many of the methods of nonrelativistic scattering
theory, methods that can be applied successfully to short-range potentials
(those that fall off with radial distance, $r,$ faster than $1/r$),
encounter divergences when applied to the infinite-range Coulomb potential.
The first Born approximation applied to the Coulomb problem produces
the Rutherford differential scattering cross section (\ref{eq:1.6}),
while the second approximation to the wavefunction diverges for all
scattering angles \cite{Dalitz1951}. Using partial wave analysis,
the sum over the angular momentum quantum number, $l,$ does not converge.
Using formal scattering theory, the transition matrix element beween
free momentum eigenvectors diverges when the energy is on shell \cite{Dettmann1971}.
Furthermore, as pointed out by Taylor \cite{Taylor1974}, the usual
asymptotic condition is not applicable and the standard definitions
of the scattering amplitude and the phase shifts are not applicable.

Other authors have shown that a wavepacket treatment can be used to
regularize these divergences \cite{Dettmann1971,Kroger1984} but did
not use the technique of partial wave analysis. Dettmann \cite{Dettmann1971}
used wavepackets with the divergent $T$ matrix elements of formal
Coulomb scattering theory to find the accepted Rutherford differential
cross section. In Section IX we will discuss his results. Kroger and
Slobodrian \cite{Kroger1984} discretized the Coulomb problem by using
a finite basis of wavepacket state vectors described by non-overlapping
step functions in momentum magnitude multiplied by spherical harmonics.
The Hamiltonian used was approximate, being the projection of the
full Hamiltonian onto this discrete basis. They calculated one $s$-wave
element of the $S$ matrix for the Coulomb interaction, not enough
to allow comparison with our results. Dollard \cite{Dollard1964}
replaced the usual asymptotic condition by a weaker, more general
condition that includes the Coulomb case. We will not find it necessary
to use his method.

Our treatment of the nonrelativistic scattering of a wavepacket (representing
a charged, spinless particle) from a Coulomb potential uses a basis
of eigenvectors of total energy and total angular momentum. As such,
it is a combination of a wavepacket treatment and partial wave analysis.
We find that the modified sum over $l$ converges to a result finite
everywhere.

Boris \textit{et al. }\cite{Boris1993}\textit{ }used both a wavepacket
treatment and partial wave analysis to give a spatial visualization
of the scattering process. In Section IX we will discuss their results.

Our method involves first finding the change of basis from free momentum
eigenvectors, $|\,\boldsymbol{k};\mathrm{free}\,\rangle$, to free
eigenvectors of the magnitude, $k$, of the momentum and the angular
momentum quantum numbers $l$ and $m$, denoted $|\,k,l,m;\mathrm{free}\,\rangle.$
These are simply related to eigenvectors of energy, $E$, and angular
momentum by (\ref{eq:2.2.5}). This is done in Section II. Then we
change the representation of a free initial wavepacket state vector
from (\ref{eq:2.1}) to (\ref{eq:2.2}). This is not an exact result
but the leading asymptotic approximation for a small fractional momentum
spread. This is done in Section III.

Then we use the technique from partial wave analysis, as is done for
short-range potentials, of adjusting the phase of this free wavefunction
to give a representation of the incoming state vector. Our wavepacket
method involves replacing a limit as time $t\rightarrow-\infty$ with
a limit as the momentum spread $\sigma_{p}\rightarrow0^{+},$ to define
the correspondence between free and incoming state vectors. This is
done in Section IV. We will discuss the role of the logarithmic phase
in (\ref{eq:4.1-1}).

We use the antiunitary time reversal operator to construct the outgoing
state vectors. The measurement is modelled as the projection onto
these outgoing state vectors.

In Section V we derive the connection, for Gaussian wavepackets, between
finite probabilities and the differential cross section.

In Section VI we evaluate an integral over momentum magnitude $k$
and numerically evaluate a sum over the angular momentum quantum number
$l$ to find the finite amplitude for the process. We explore the
parameter space and present our results in Section VII, with consideration
of the possibility of experimental measurement of the phenomena we
predict.

In Section VIII we present an integral relation satisfied by our scattering
probability that represents the conservation of total probability.
We use this as a check on our results. We also discuss the optical
theorem in the context of Coulomb wavepacket scattering.

In Section IX we compare our results with those of other authors.
Conclusions are presented in Section X.

For the nonrelativistic scattering of a spinless, charged particle
in a static Coulomb field, the time-independent Schrödinger equation,
with Hamiltonian in the $\boldsymbol{x}$ representation
\begin{equation}
H=-\frac{1}{2m_{0}}\nabla^{2}+\frac{Z_{1}Z_{2}\alpha}{|\boldsymbol{r}|},\label{eq:1.0}
\end{equation}
can be solved to find an energy eigenvector of the form
\begin{equation}
\psi(\boldsymbol{x},p)=e^{ipz}f(r-z),\label{eq:1.0.5}
\end{equation}
as done in standard textbooks \cite{Messiah1961,Merzbacher1998}.
The asymptotic form of this wavefunction as $|r-z|\rightarrow\infty$
then gives the scattering amplitude
\begin{equation}
f(\theta,p)=-\frac{\eta}{2p\sin^{2}(\frac{\theta}{2})}\exp(-i\eta\ln(\sin^{2}(\frac{\theta}{2}))+i2\sigma_{0}).\label{eq:1.1}
\end{equation}
(Throughout this paper, we use Heaviside-Lorentz units, in which $\hbar=c=\epsilon_{0}=\mu_{0}=1$.)
The momentum magnitude, $p,$ in terms of the energy, $E,$ is $p=\sqrt{2m_{0}E}.$
The scattered wave is observed at the scattering angle $\theta$ to
the incident direction. The dimensionless quantity $\eta$ is
\begin{equation}
\eta=Z_{1}Z_{2}\frac{\alpha}{\beta},\label{eq:1.2}
\end{equation}
where $Z_{1}$ is the atomic number of the Coulomb field, $Z_{2}$
is the atomic number of the incident particle, and $\beta=p/m_{0}$
is the nonrelativistic form of the speed of the incoming particle
of mass $m_{0}$. The quantity $\alpha$ is the fine structure constant
$\alpha=e^{2}/4\pi\cong1/137$. The phase shifts for general $l$
are defined by (\ref{eq:4.2-1}).

To make contact with experiment, the Rutherford differential cross
section is then given by
\begin{eqnarray}
\frac{d\sigma}{d\Omega} & = & |f(\theta,p)|^{2}\label{eq:1.5}\\
 & = & \frac{Z_{1}^{2}Z_{2}^{2}\alpha^{2}}{16E^{2}\sin^{4}(\frac{\theta}{2})}.\label{eq:1.6}
\end{eqnarray}
Rutherford used an entirely classical treatment of Coulomb scattering
of point particles to obtain, remarkably, the same result as the later
quantum calculation \cite{Rutherford1911}. The experimental verification
of this cross section formula is routinely done as a laboratory activity
in undergraduate physics courses.

This scattering amplitude (\ref{eq:1.1}) is not an amplitude of the
type for which the modulus-squared is a probability less than or equal
to unity. For a start it is not dimensionless, instead having the
dimensions of length. Furthermore it diverges in the forward direction,
$\theta=0$, for all momenta, and as $p\rightarrow0^{+}$ for all
scattering angles.

One purpose of this paper is to find the relation between the scattering
amplitude and what we will call the \textit{finite amplitude} for
this scattering process, which is simply the $S$ matrix element between
normalized wavepacket state vectors, and as such must be less than
or equal to unity in magnitude. This relation, which we will derive
for Gaussian wavepackets in Section V, will, in the case of more general
interactions, only hold for scattering angles sufficiently far from
the forward direction.

\section{The change of improper basis vectors}

We first need to change the representation of a \textit{free} wavepacket
state vector, normalized to unity, from
\begin{equation}
|\,\psi;\mathrm{free}\,\rangle=\int d^{3}k\,|\,\boldsymbol{k};\mathrm{free}\,\rangle\psi_{\mathrm{free}}(\boldsymbol{k})\label{eq:2.1}
\end{equation}
to
\begin{equation}
|\,\psi;\mathrm{free}\,\rangle=\int_{0}^{\infty}dk\sum_{l=0}^{\infty}\sum_{m=-l}^{l}|\,k,l,m;\mathrm{free}\,\rangle\Psi_{\mathrm{free}}(k,l,m),\label{eq:2.2}
\end{equation}
where the $|\,k,l,m;\mathrm{free}\,\rangle$ are eigenvectors of the
magnitude of momentum ($k$) and angular momentum (the familiar quantum
numbers $l$ and $m$ taking only integer values). Note that eigenvectors
of energy ($E$) and angular momentum are simply related to these
by a change of normalization:
\begin{equation}
|\,E,j,m\,\rangle=|\,k,j,m\,\rangle\sqrt{\frac{m_{0}}{k}}.\label{eq:2.2.5}
\end{equation}

We start with the improper basis vectors. The basis transformation
is derived by Sakurai \cite{Sakurai1994} and agrees with our calculation.
We want the $|\,\boldsymbol{k};\mathrm{free}\,\rangle$ to have the
orthonormality relation
\begin{equation}
\langle\,\boldsymbol{k}_{1};\mathrm{free}\,|\,\boldsymbol{k}_{2};\mathrm{free}\,\rangle=\delta^{3}(\boldsymbol{k}_{1}-\boldsymbol{k}_{2})\label{eq:2.10}
\end{equation}
and the $|\,k,l,m\,\rangle$ to obey the orthonormality relation
\begin{equation}
\langle\,k_{1},l_{1},m_{1};\mathrm{free}\,|\,k_{2},l_{2},m_{2};\mathrm{free}\,\rangle=\delta(k_{1}-k_{2})\delta_{l_{1}l_{2}}\delta_{m_{1}m_{2}}.\label{eq:2.11-1}
\end{equation}
The result is
\begin{equation}
|\,\boldsymbol{k};\mathrm{free}\,\rangle=\sum_{l=0}^{\infty}\sum_{m=-l}^{l}|\,k,l,m;\mathrm{free}\,\rangle Y_{lm}^{*}(\hat{\boldsymbol{k}})\frac{1}{k},\label{eq:2.11}
\end{equation}
with $k=|\boldsymbol{k}|$, with a choice of global phase.

\section{Basis transformation for a free wavepacket state vector}

The free state vector that will be modified to become the incoming
state vector is chosen as the Gaussian wavepacket Schrödinger picture
state vector
\begin{equation}
|\,\boldsymbol{p}_{i},\boldsymbol{R}_{i};\sigma_{p};\mathrm{free}\,(0)\rangle=\int d^{3}k\,|\,\boldsymbol{k};\mathrm{free}\,\rangle e^{-i\boldsymbol{k}\cdot\boldsymbol{R}_{i}}\frac{e^{-|\boldsymbol{k}-\boldsymbol{p}_{i}|^{2}/4\sigma_{p}^{2}}}{(2\pi\sigma_{p}^{2})^{\frac{3}{4}}}.\label{eq:3.1}
\end{equation}
The average momentum is $\boldsymbol{p}_{i}=p\hat{\boldsymbol{z}.}$
The standard deviation of momentum in all directions is $\sigma_{p}$.
We are going to choose a well-resolved momentum,
\begin{equation}
\epsilon=\frac{\sigma_{p}}{p}\ll1,\label{eq:3.3}
\end{equation}
and we will only need the leading asymptotic approximation to the
wavefunction $\Psi_{\mathrm{free}}(k,j,m)$ in powers of $\sqrt{\epsilon}$,
as will be explained shortly. The average position at $t=0$ is $\boldsymbol{R}_{i}=-R\hat{\boldsymbol{z}}.$
The standard deviation of position in all directions at $t=0$ is
$\sigma_{x}$ with $\sigma_{x}\sigma_{p}=1/2.$

The distance, $R$, from the center of the Coulomb potential at $t=0$
must be large so that the corresponding \textit{interacting} state
vector is effectively free at that time, and so that we can apply
an approximation in Section~V. However there must be a limit on the
size of $R.$ In our scattering geometry, over the time of the experiment,
the incoming wavepacket will traverse a distance of about $2R.$ We
demand that the \textit{spreading} of the wavepacket be negligible
over this time. The nonrelativistic law for spreading of Gaussian
wavepackets is \cite{Merzbacher1998} 
\begin{equation}
\sigma_{x}(t)=\sqrt{\sigma_{x}^{2}+(\frac{\sigma_{p}}{p})^{2}(\beta t)^{2}},\label{eq:3.6}
\end{equation}
where $t$ is the time measured from minimality. We see that spreading
becomes significant for $\beta t\sim(p/\sigma_{p})\sigma_{x}.$ If
we instead choose a particular dependence of $R$ on $p$ and $\sigma_{p},$
\begin{eqnarray}
R & = & \sqrt{\frac{p}{\sigma_{p}}}\sigma_{x}=\frac{1}{\sqrt{\epsilon}}\sigma_{x},\label{eq:3.9}
\end{eqnarray}
we will still have a very large initial distance for $\epsilon\ll1$,
much larger that the initial spread, but wavepacket spreading will
be negligible over the time of the experiment. We note that the dependence
on $R$ of the probability that we calculate (\ref{eq:7.8}) will
only influence the time shifts, and with only a logarithmic dependence.
For short-range potentials, the absence of the logarithmic phase will
mean that the time shifts are independent of $R.$

Now we transform the wavefunction. Using (\ref{eq:2.1}),(\ref{eq:2.2})
and (\ref{eq:2.11}), we find the transformation
\begin{equation}
\Psi_{\mathrm{free}}(k,l,m)=k\int_{0}^{\pi}\sin\theta_{k}d\theta_{k}\int_{0}^{2\pi}d\varphi_{k}\,Y_{lm}^{*}(\theta_{k},\varphi_{k})\psi_{\mathrm{free}}(\boldsymbol{k}),\label{eq:3.10.1}
\end{equation}
again with $k=|\boldsymbol{k}|$ and with $\hat{\boldsymbol{k}}=(\theta_{k},\varphi_{k})$.

The spherical harmonic is
\begin{eqnarray}
Y_{lm}^{*}(\hat{\boldsymbol{k}}) & = & \sqrt{\frac{2l+1}{4\pi}}e^{-im\varphi_{k}}d_{m0}^{l}(\theta_{k}),\label{eq:3.11}
\end{eqnarray}
where
\begin{equation}
d_{m_{1}m_{2}}^{l}(\theta_{k})=\langle\,l,m_{1}\,|\,e^{-i\theta_{k}J_{y}}\,|\,l,m_{2}\,\rangle\label{eq:3.12}
\end{equation}
are Wigner rotation matrices \cite{Messiah1961}.

The $\varphi_{k}$ integral is simply
\begin{equation}
\int_{0}^{2\pi}d\varphi_{k}\,e^{-im\varphi_{k}}=2\pi\,\delta_{m0}.\label{eq:3.14}
\end{equation}

We have
\begin{equation}
|\boldsymbol{k}-\boldsymbol{p}_{i}|^{2}=(k-p)^{2}+2kp(1-\cos\theta_{k}).\label{eq:3.14.1}
\end{equation}
To eventually calculate an amplitude for the scattering process, we
will be evaluating an integral over $k$ in which a factor $g(k)^{2}$
will appear, where
\begin{equation}
g(k)=e^{-(k-p)^{2}/4\sigma_{p}^{2}}.\label{eq:3.15-1}
\end{equation}
This factor is sharply peaked in $k$ with a width of $\mathcal{O}(\sigma_{p})$.
We will repeatedly find integrands similarly dominated by narrow factors,
including the integral over $\theta_{k}$, which contains the factor
\begin{equation}
h(\theta_{k})=e^{-kp(1-\cos\theta_{k})/2\sigma_{p}^{2}}.\label{eq:3.16}
\end{equation}
This function is sharply peaked at $\theta_{k}=0$ with a width of
order $\sigma_{p}/p.$ We need to find the leading approximation to
such integrals for $\sigma_{p}/p\ll1,$ along with an estimate of
the remainder. We use the technique of expanding the exponents (and
more slowly varying factors) in powers of $k-p,$ treating this as
a quantity of $\mathcal{O}(\sigma_{p}),$ and in powers of $\theta_{k},$
treating this as a quantity of $\mathcal{O}(\sigma_{p}/p).$ We then
evaluate the integrals to lowest order and estimate that the fractional
remainder is no larger than $\mathcal{O}(\sqrt{\epsilon}),$ uniform
in the remaining variables.

With this method, we find
\begin{equation}
\frac{kp}{\sigma_{p}^{2}}(1-\cos\theta_{k})=\frac{p^{2}\theta_{k}^{2}}{2\sigma_{p}^{2}}+\mathcal{O}(\epsilon).\label{eq:3.17}
\end{equation}
Similarly, we find that we can replace
\begin{eqnarray}
-i\boldsymbol{k}\cdot\boldsymbol{R}_{i} & = & +ikR+\mathcal{O}(\sqrt{\epsilon})\label{eq:3.18}
\end{eqnarray}
using our scheme to avoid wavepacket spreading. Here $k$ cannot be
replaced by $p.$

With these approximations, we need to evaluate the integral (after
using $\sin\theta_{k}=\theta_{k}(1+\mathcal{O}(\epsilon^{2}))$)
\begin{equation}
I(l,\epsilon)=\int_{0}^{\pi}\theta_{k}\,d\theta_{k}\,e^{-p^{2}\theta_{k}^{2}/4\sigma_{p}^{2}}\,d_{00}^{l}(\theta_{k}).\label{eq:3.19}
\end{equation}
We cannot use a power series in $\theta_{k}$ for the Wigner rotation
matrix, $d_{00}^{l}(\theta_{k})$, as for large $l$ it will have
rapid variation on the region $0\leq\theta_{k}\apprle\epsilon$. Instead,
we use an approximation, from a paper by the author in preparation,
valid for small $\theta,$ uniform in $l$:
\begin{equation}
d_{00}^{l}(\theta_{k})=J_{0}(\sqrt{l(l+1)+\frac{1}{3}}\,\theta_{k})\{1+\mathcal{O}(\theta_{k}^{2})\}\label{eq:3.20}
\end{equation}
(This result has been checked numerically.) Then the integral evaluates
to \cite{Gradsteyn1980}
\begin{equation}
I(l,\epsilon)=\frac{2\sigma_{p}^{2}}{p^{2}}\exp(-\sigma_{p}^{2}(l+\frac{1}{2})^{2}/p^{2})\{1+\mathcal{O}(\epsilon^{2})\}.\label{eq:3.21}
\end{equation}

Finally the leading asymptotic approximation to the $k,j,m$ representation
of the free state vector is given by the wavefunction (normalized
up to $\mathcal{O}(\epsilon^{2})$ corrections)
\begin{equation}
\Psi_{\mathrm{free}}(k,j,m\,|\,\boldsymbol{p}_{i},\boldsymbol{R}_{i};\sigma_{p})=\delta_{m0}\,e^{ikR}\,\frac{e^{-(k-p)^{2}/4\sigma_{p}^{2}}}{(2\pi\sigma_{p}^{2})^{\frac{1}{4}}}\frac{\sqrt{l+\frac{1}{2}}\,e^{-\sigma_{p}^{2}(l+\frac{1}{2})^{2}/p^{2}}}{p/2\sigma_{p}}.\label{eq:3.22}
\end{equation}
Note that a narrow distribution in angle produces a wide distribution
in angular momentum.

\section{Construction of the incoming and outgoing state vectors}

We will be considering a range of different incident momentum magnitudes,
$p$. In each case, we require that the fractional momentum spread,
$\sigma_{p}(p)/p$, be small. The simplest scheme is to make this
quantity a constant independent of momentum:
\begin{equation}
\frac{\sigma_{p}(p)}{p}\equiv\epsilon\ll1.\label{eq:4.5-1}
\end{equation}
It will also become necessary to impose a lower limit on momentum
for a given $\epsilon$, thereby imposing an upper limit on the quantity
$|\eta(p)|$, in (\ref{eq:4.26.2}) below. Note that we can still
consider arbitrarily small momentum magnitudes, but only by decreasing
$\epsilon$. In our numerical calculations, we will use $\epsilon=0.001,$
which gives $\sqrt{\epsilon}\cong0.032$, sufficiently small for our
purposes.

In scattering theory \cite{Newton1982,Taylor1974}, we define two
mappings from free state vectors to interacting state vectors, provided
by the Møller operators $\Omega^{(+)}$ and $\Omega^{(-)}.$ It is
supposed that for an arbitrary free state vector, an interacting state
vector can always be constructed that behaves in an essentially free
manner at very early times, with properties matching those of the
free state vector at those early times. We say that this interacting
state vector is the incoming state vector corresponding to that free
state vector, and write
\begin{equation}
|\,\mathrm{incoming}\,\rangle=\Omega^{(+)}|\,\mathrm{free}\,\rangle.\label{eq:4.8}
\end{equation}
It should be clear that this correspondence only makes physical sense
for wavepacket state vectors, where there is a position distribution
that may be localized far from the scattering center at very early
times. However, once it is constructed, the Møller operator will have
matrix elements between plane-wave improper state vectors.

The other Møller operator, $\Omega^{(-)},$ maps from free to outgoing
state vectors, such that the latter are essentially free as $t\rightarrow+\infty.$

The incoming correspondence is defined by requiring
\begin{equation}
\lim_{t\rightarrow-\infty}\int d^{3}r\,|\psi_{\mathrm{in}}(\boldsymbol{r},t)-\psi_{\mathrm{free}}(\boldsymbol{r},t)|^{2}=0.\label{eq:4.9}
\end{equation}

In contrast, in our method we only need to find the free to interacting
correspondence for two particular sets of state vectors, namely
\begin{equation}
|\,\boldsymbol{p}_{i},\boldsymbol{R}_{i},\sigma_{p};\mathrm{in}\,\rangle=\Omega^{(+)}|\,\boldsymbol{p}_{i},\boldsymbol{R}_{i},\sigma_{p};\mathrm{free}\,\rangle\label{eq:4.10}
\end{equation}
and
\begin{equation}
|\,\boldsymbol{p}_{f},\boldsymbol{R}_{f},\sigma_{p};\mathrm{out}\,\rangle=\Omega^{(-)}|\,\boldsymbol{p}_{f},\boldsymbol{R}_{f},\sigma_{p};\mathrm{free}\,\rangle,\label{eq:4.11}
\end{equation}
where the latter will be defined below. We consider the incoming case
first.

Since, for small $\epsilon$, the free wavepacket is already localized
far from the scattering centre at $t=0,$ we argue that the interacting
state vector should be essentially free at $t=0,$ not just as $t\rightarrow-\infty.$
This leads to two constraints. The probability distribution in $k,l$
and $m$ is independent of time for the interacting system, so it
must be equal to its value as $t\rightarrow-\infty,$ when we argue
that it must take its free value exactly (for negligible bound state
content in the attractive case). Hence
\begin{equation}
|\Psi_{\mathrm{in}}(k,l,m)|^{2}=|\Psi_{\mathrm{free}}(k,l,m)|^{2}.\label{eq:4.12}
\end{equation}
(The modifications imposed by the logarithmic phase in (\ref{eq:4.1-1})
will not negate this argument.) In practice we equate $|\Psi_{\mathrm{in}}(k,l,m)|^{2}$
to the leading asymptotic approximation as $\epsilon\rightarrow0^{+}$
of $|\Psi_{\mathrm{free}}(k,l,m)|^{2},$ given by (\ref{eq:3.22}),
with the knowledge that the error is negligible for the value of $\epsilon$
that we consider.

We will find that it is not possible for the phases as well as the
magnitudes of the incoming and free position probability amplitudes
to be equal in the limit (\ref{eq:4.9}). Instead we propose that
it is sufficient to require that the position probability density
at $t=0$ approach its free form in the limit as $\epsilon\rightarrow0^{+},$
\begin{equation}
\lim_{\epsilon\rightarrow0^{+}}\{|\psi_{\mathrm{in}}(\boldsymbol{r},0)|^{2}-|\psi_{\mathrm{free}}(\boldsymbol{r},0)|^{2}\}=0.\label{eq:4.13}
\end{equation}
(This argument will need to be modified when taking into account the
effect of the logarithmic phase.) Compared to (\ref{eq:4.9}), there
is no need here to evolve the wavefunctions towards infinite times.

The first of these equations implies that the interacting and free
$k,l,m$ wavefunctions can only differ by a phase factor
\begin{equation}
\Psi_{\mathrm{in}}(k,l,m)=e^{i\chi_{l}(k)}\Psi_{\mathrm{free}}(k,l,m).\label{eq:4.14}
\end{equation}
The fact that $\chi_{l}(k)$ is independent of $m$ follows from the
extra physical requirement that rotations commute with the Møller
operators. (The phase factor $\exp(ikR)$ from $\Psi_{\mathrm{free}}(k,j,m\,|\,\boldsymbol{p}_{i},\boldsymbol{R}_{i};\sigma_{p})$
is physically relevant and will be retained.)

Now we impose the constraint (\ref{eq:4.13}) on the incoming position
probability amplitude,
\begin{equation}
\psi_{\mathrm{in}}(\boldsymbol{r},0)=\int_{0}^{\infty}dk\sum_{l=0}^{\infty}\sum_{m=-l}^{l}\langle\,\boldsymbol{r}\,|\,k,l,m\,\rangle e^{i\chi_{l}(k)}e^{ikR}|\Psi_{\mathrm{free}}(k,l,m)|.\label{eq:4.15}
\end{equation}
It is clear that for large $R$ we can use an approximation for large
$r$ if the incoming probability density is to have a similar form
to the free density. We use the known solutions \cite{Messiah1961}
(with continuum normalization similar to (\ref{eq:2.11-1})) with
asymptotic form as $r=|\boldsymbol{r}|\rightarrow\infty,$
\begin{equation}
\langle\,\boldsymbol{r}\,|\,k,l,m\,\rangle\sim\sqrt{\frac{2}{\pi}}\,Y_{lm}(\hat{\boldsymbol{r}})\,\frac{1}{r}\sin(kr-\eta(k)\ln(2kr)-l\frac{\pi}{2}+\sigma_{l}(k)),\label{eq:4.1-1}
\end{equation}
where the Coulomb phase shifts, $\sigma_{l}$, are defined implicitly
by
\begin{equation}
e^{i2\sigma_{l}(k)}=\frac{\Gamma(l+1+i\eta(k))}{\Gamma(l+1-i\eta(k))}.\label{eq:4.2-1}
\end{equation}
In contrast, the free spherical waves have the asymptotic form
\begin{equation}
\langle\,\boldsymbol{r}\,|\,k,l,m;\mathrm{free}\,\rangle\sim\sqrt{\frac{2}{\pi}}\,Y_{lm}(\hat{\boldsymbol{r}})\,\frac{1}{r}\sin(kr-l\frac{\pi}{2})\label{eq:4.3-1}
\end{equation}
and the free position probability amplitude is
\begin{equation}
\psi_{\mathrm{free}}(\boldsymbol{r},0)=\int_{0}^{\infty}dk\sum_{l=0}^{\infty}\sum_{m=-l}^{l}\langle\,\boldsymbol{r}\,|\,k,l,m;\mathrm{free}\,\rangle e^{ikR}|\Psi_{\mathrm{free}}(k,l,m)|,\label{eq:416}
\end{equation}
which evaluates to
\begin{equation}
\psi_{\mathrm{free}}(\boldsymbol{r},0)=e^{i\boldsymbol{p}_{i}\cdot(\boldsymbol{r}-\boldsymbol{R}_{i})}\frac{e^{-|\boldsymbol{r}-\boldsymbol{R}_{i}|^{2}/4\sigma_{x}^{2}}}{(2\pi\sigma_{x}^{2})^{\frac{3}{4}}}.\label{eq:4.17}
\end{equation}

Note that for short-range potentials, there will be different phase
shifts instead of $\sigma_{l}(k)$ and the logarithmic phase term
$-\eta(k)\ln(2kr)$ will be absent.

In
\begin{equation}
\sin\varphi=\frac{1}{2i}(e^{+i\varphi}-e^{-i\varphi})\label{eq:4.18}
\end{equation}
with
\begin{equation}
\varphi=kr-\eta(k)\ln(2kr)-l\frac{\pi}{2}+\sigma_{l}(k),\label{eq:4.19}
\end{equation}
only the integral over $k$ containing $\exp(-i\varphi)\exp(+ikR)$
will be significant on $r\geq0.$ So the relevant phase to consider
is
\begin{equation}
\Phi(k,r)=kR-kr+\eta(k)\ln(2kr)-\sigma_{l}(k)+\chi_{l}(k),\label{eq:4.20}
\end{equation}
in comparison with
\begin{equation}
\Phi_{\mathrm{free}}(k,r)=kR-kr,\label{eq:4.21}
\end{equation}
which leads to a peak probability density at $r=R.$

We would like to be able to choose $\chi_{l}(k)$ to cancel both the
Coulomb phase shifts and the logarithmic phase. This is not possible,
since the latter is a function of $r$ and $\chi_{l}(k)$ cannot depend
on $r.$ Instead, we choose
\begin{equation}
\chi_{l}(k)=\sigma_{l}(k)\label{eq:4.22}
\end{equation}
and investigate the physical effects of the logarithmic phase.

We expand in powers of $k-p$:
\begin{equation}
\eta(k)\ln(2kr)\cong\eta(p)\ln(2pr)-\frac{(k-p)}{p}\eta(p)(\ln(2pr)-1).\label{eq:4.23}
\end{equation}
The remainder is found to be negligible for $r=R$ and $|\eta(p)|=844$
(see (\ref{eq:4.26.2}) below).

The first term in (\ref{eq:4.23}) is independent of $k$ so merely
contributes an $r$-dependent phase factor to the position wavefunction.
The second term produces a spatial shift of the peak of the position
wavefunction away from its free value of $R.$ To see this, we note
that $\Phi(k,r)$ is stationary in $k$ at the peak, $k=p,$ for $r=\bar{r}$
satisfying
\begin{equation}
R-\bar{r}-\frac{\eta(p)}{p}(\ln(2p\bar{r})-1)=0.\label{eq:4.24}
\end{equation}
Inverting for large $R$ gives the asymptotic behaviour
\begin{equation}
\bar{r}(R)\sim R-\Delta(R),\label{eq:4.25}
\end{equation}
with
\begin{equation}
\Delta(R)=\frac{\eta(p)}{p}(\ln(2pR)-1).\label{eq:4.26}
\end{equation}
(It will be easily seen that the same result will hold for the outgoing
state vectors.)

We suppose that $|\Delta(R)|$ could be as large as $R/2$ without
affecting the negligibility of wavepacket spreading or the requirement
that the mean initial radial distance of the wavepacket grow as $\epsilon^{-3/2}$
for given momentum. This leads to
\begin{equation}
|\eta(p)|\leq\frac{1}{4\epsilon^{\frac{3}{2}}|\frac{3}{2}\ln\frac{1}{\epsilon}-1|}=844\label{eq:4.26.2}
\end{equation}
for $\epsilon=0.001$.

So, with our choice of phase (\ref{eq:4.22}) to define the incoming
state vector corresponding to a free state at average radial position
$R,$ the actual average position of the incoming state vector is
at radius $R-\Delta(R).$ It is clear that the incoming state can
never be considered entirely free, since evolution over short times
will always give
\begin{equation}
U(t)\,|\,\boldsymbol{p}_{i},\boldsymbol{R}_{i},\sigma_{p};\mathrm{in}\,\rangle\neq|\,\boldsymbol{p}_{i},\boldsymbol{R}_{i}+\frac{\boldsymbol{p}_{i}}{m_{0}}t,\sigma_{p};\mathrm{in}\,\rangle.\label{eq:4.27}
\end{equation}

We argue that this is the correct physical behaviour for a particle
in a Coulomb field because very similar behaviour is seen in the classical
case at large distances from the origin. A straightforward calculation
of the classical trajectories for a head-on collision shows that if
free motion is defined as
\begin{equation}
R(t)=R_{0}+\frac{p}{m_{0}}t,\label{eq:4.28}
\end{equation}
where $p$ is the finite limit of momentum magnitude far from the
origin, then the actual position of the particle has the asymptotic
form for large $R$
\begin{equation}
\bar{r}(R)\sim R-\frac{\eta(p)}{p}(\ln(2pR)-\ln\eta),\label{eq:4.29}
\end{equation}
in asymptotic agreement with the quantum calculation as $R\rightarrow\infty$.

A longitudinal spatial shift will not change the $k,l,m$ probability
density, so the conclusion (\ref{eq:4.12}) stands. We now see how
the constraint (\ref{eq:4.13}) must be modified to
\begin{equation}
\lim_{\epsilon\rightarrow0^{+}}\{|\psi_{\mathrm{in}}(\boldsymbol{r},0)|^{2}-|\psi_{\mathrm{free}}(\boldsymbol{r}-\Delta(R)\hat{\boldsymbol{p}}_{i},0)|^{2}\}=0.\label{eq:4.31}
\end{equation}
Clearly this result is satisfied in our case.

Our final result for the wavefunction of the incoming state vector
is
\begin{equation}
\Psi_{\mathrm{in}}(k,l,m\,|\,\boldsymbol{p}_{i},\boldsymbol{R}_{i};\sigma_{p})=e^{+i\sigma_{l}(k)}e^{+ikR}\delta_{m0}|\Psi_{\mathrm{free}}(k,l,0\,|\,\boldsymbol{p}_{i},\boldsymbol{R}_{i};\sigma_{p})|.\label{eq:4.32}
\end{equation}

We use the antiunitary time reversal operator, $A(\mathcal{T}),$
to construct the outgoing state vector as the time reversal of an
incoming state vector is an outgoing state vector. It can be shown
\cite{Merzbacher1998} that the transformation law of the interacting
eigenvectors is
\begin{equation}
A(\mathcal{T})\,|\,k,l,m\,\rangle=(-)^{l+m}|\,k,l,-m\,\rangle.\label{eq:4.20-1}
\end{equation}
Starting with
\begin{equation}
|\,\boldsymbol{p}_{i},\boldsymbol{R}_{i},\sigma_{p};\mathrm{in}\,\rangle=\int_{0}^{\infty}dk\sum_{l=0}^{\infty}|\,k,l,0\,\rangle\Psi_{\mathrm{in}}(k,l,0)\label{eq:4.21-1}
\end{equation}
and applying a time reversal, a rotation by $\pi$ about the $y$
axis and then a further rotation by $\theta$ about the $y$ axis
gives
\begin{equation}
|\,\boldsymbol{p}_{f},\boldsymbol{R}_{f},\sigma_{p};\mathrm{out}\,\rangle\equiv U(R_{y}(\pi+\theta))A(\mathcal{T})\,|\,\boldsymbol{p}_{i},\boldsymbol{R}_{i},\sigma_{p};\mathrm{in}\,\rangle,\label{eq:4.22-1}
\end{equation}
so
\begin{equation}
|\,\boldsymbol{p}_{f},\boldsymbol{R}_{f},\sigma_{p};\mathrm{out}\,\rangle=\int_{0}^{\infty}dk\sum_{l=0}^{\infty}\sum_{m=-l}^{l}|\,k,l,m\,\rangle\,d_{m0}^{l}(\theta)\,\Psi_{\mathrm{in}}^{*}(k,l,0).\label{eq:4.23-1}
\end{equation}
We have chosen
\begin{equation}
\boldsymbol{R}_{f}=R\,(\sin\theta,0,\cos\theta).\label{eq:4.24-1}
\end{equation}
So
\begin{equation}
\Psi_{\mathrm{out}}(k,l,m\,|\,\boldsymbol{p}_{f},\boldsymbol{R}_{f};\sigma_{p})=e^{-i\sigma_{l}(k)}e^{-ikR}\,d_{m0}^{l}(\theta)|\Psi_{\mathrm{free}}(k,l,0\,|\,\boldsymbol{p}_{i},\boldsymbol{R}_{i};\sigma_{p})|.\label{eq:4.33}
\end{equation}
Calculation of $\psi_{\mathrm{out}}(r,0)$ involves the $\exp(+i\varphi)$
factor and arrives at the same result (\ref{eq:4.25},\ref{eq:4.26})
for the shift due to the logarithmic phase term.

Our scattering geometry is shown in Figure 2.

\begin{figure}
\begin{centering}
\includegraphics[width=8.6cm]{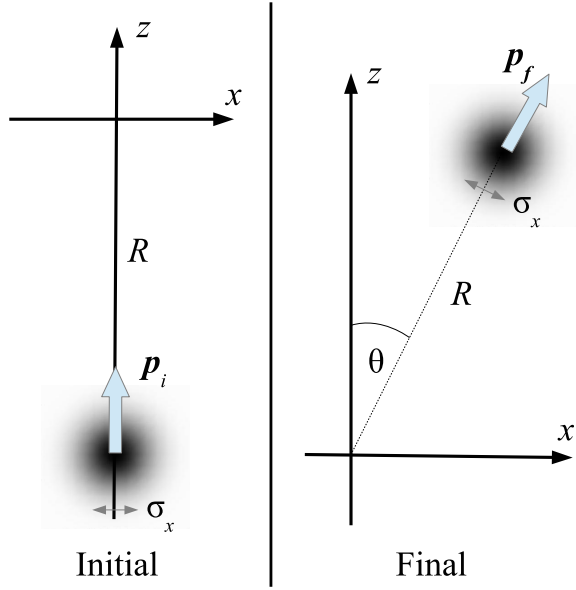}
\par\end{centering}
\caption{Scattering geometry.}

\end{figure}

\section{Connection between finite amplitudes, scattering amplitudes and $S$
matrix elements}

From the way we have defined the incoming and outgoing state vectors,
as illustrated in Figure 1, we see that we need to apply a time shift
of $T\cong2R/\beta$ (for low interaction strength $\eta$) to the
incoming state vector to align the wavepackets and maximize the scattering
probability. Then both wavepackets are expected to be localized around
the origin at time $t\cong-R/\beta.$ We leave $T$ as a variable,
a parameter of the incoming state vector. Later we will scan in $T$
(or rather $\delta$ as defined in (\ref{eq:7.6.7},\ref{eq:7.6.8})
below) to see the time delays or advancements away from free evolution
caused by the interaction. So we have the incoming and outgoing state
vectors
\begin{equation}
|\,i\,\rangle=\hat{\Omega}^{(+)}\{e^{-i\hat{H}_{0}T}\,|\,\boldsymbol{p}_{i},\boldsymbol{R}_{i},\sigma_{p};\mathrm{free}(0)\,\rangle\}\quad\mathrm{and}\quad|\,f\,\rangle=\hat{\Omega}^{(-)}\{|\,\boldsymbol{p}_{f},\boldsymbol{R}_{f},\sigma_{p};\mathrm{free}(0)\,\rangle\},\label{eq:5.1.1}
\end{equation}
respectively.

The quantity that we call a finite amplitude is just an $S$ matrix
element between normalized free wavepacket state vectors
\begin{equation}
\langle\,f\,|\,i\,\rangle=\langle\,\boldsymbol{p}_{f},\boldsymbol{R}_{f},\sigma_{p};\mathrm{free}(0)\,|\,\hat{S}\,|\,\boldsymbol{p}_{i},\boldsymbol{R}_{i},\sigma_{p};\mathrm{free}(T)\,\rangle.\label{eq:5.0}
\end{equation}
We used the definition of the $S$ matrix \cite{Newton1982}
\begin{equation}
\hat{S}=\hat{\Omega}^{(-)\dagger}\hat{\Omega}^{(+)}.\label{eq:47.2}
\end{equation}
Note that $S$ matrix elements (and thus the cross section) are independent
of Schrödinger picture evolution time, $t,$
\begin{equation}
\langle\,\boldsymbol{p}_{f},\boldsymbol{R}_{f},\sigma_{p};\mathrm{free}(t)\,|\,\hat{S}\,|\,\boldsymbol{p}_{i},\boldsymbol{R}_{i},\sigma_{p};\mathrm{free}(T+t)\,\rangle=\langle\,\boldsymbol{p}_{f},\boldsymbol{R}_{f},\sigma_{p};\mathrm{free}(0)\,|\,\hat{S}\,|\,\boldsymbol{p}_{i},\boldsymbol{R}_{i},\sigma_{p};\mathrm{free}(T)\,\rangle,\label{eq:47.2.5}
\end{equation}
since the $S$ matrix commutes with the full Hamiltonian. This, in
turn, follows from the fact that the action of the Møller operators
is to simply multiply $\Psi_{\mathrm{free}}(k,l,m)$ by phase factors
that depend only on $k$ and $l$.

For state vectors normalized to unity, the Schwartz inequality \cite{Messiah1961}
guarantees
\begin{equation}
|\langle\,f\,|\,i\,\rangle|^{2}\leq1,\label{eq:47.3}
\end{equation}
which is why we call these ``finite amplitudes''.

$S$ matrix elements between momentum eigenvectors are related to
scattering amplitudes by (see Wichmann \cite{Wichmann1965} for a
derivation using wavepackets)

\begin{equation}
\langle\,\boldsymbol{k}_{f};\mathrm{free}\,|\,\hat{S}\,|\,\boldsymbol{k}_{i};\mathrm{free}\,\rangle=\delta^{3}(\boldsymbol{k}_{f}-\boldsymbol{k}_{i})+\frac{i}{2\pi k_{i}}\delta(k_{f}-k_{i})\,f(\theta,k_{i}).\label{eq:5.1}
\end{equation}
With $\hat{S}=1+i\hat{T},$ we can find the $\hat{T}$ matrix element
between our Gaussian wavepacket state vectors, by approximating the
integrals for $\sigma_{p}/p\ll1$, to give
\begin{equation}
\langle\,\boldsymbol{p}_{f},\boldsymbol{0};\sigma_{p};\mathrm{free}\,|\,\hat{T}\,|\,\boldsymbol{p}_{i},\boldsymbol{0};\sigma_{p};\mathrm{free}\,\rangle=\frac{4\sigma_{p}^{2}}{p}f(\theta,p)\{1+\mathcal{O}(\epsilon)\}.\label{eq:5.2}
\end{equation}
(The average positions of the wavepackets are not important to this
calculation, but will be relevant later.)

Then we find the differential cross section to be (using (\ref{eq:1.5})),
for $\epsilon\ll1$,
\begin{equation}
\frac{d\sigma}{d\Omega}|_{\mathrm{S-R}}=\frac{p^{2}}{16\sigma_{p}^{4}}|\langle\,\boldsymbol{p}_{f},\boldsymbol{0};\sigma_{p};\mathrm{free}\,|\,\hat{T}\,|\,\boldsymbol{p}_{i},\boldsymbol{0};\sigma_{p};\mathrm{free}\,\rangle|^{2}.\label{eq:5.3}
\end{equation}

In Section VIII we will discuss how the separation $\hat{S}=1+i\hat{T}$
of the $S$ matrix gives physically relevant results for short-range
potentials, where the probability is always found to have a peak in
the forward direction of height slightly less than unity and width
$\Delta\theta\sim\epsilon$ (produced by the 1 term) and scattering
in other directions at a much smaller probability (produced by the
$\hat{T}$ term). Then the expression (\ref{eq:5.3}) is appropriate
for short-range ($S-R$) potentials. We will find for Coulomb scattering
that this separation does not produce physically meaningful terms,
and that (except at very low interaction strengths/high energies)
there is no narrow forward peak. Then it is more appropriate to combine
the terms and write
\begin{equation}
\frac{d\sigma}{d\Omega}|_{\mathrm{C}}=\frac{p^{2}}{16\sigma_{p}^{4}}|\langle\,\boldsymbol{p}_{f},\boldsymbol{0};\sigma_{p};\mathrm{free}\,|\,\hat{S}\,|\,\boldsymbol{p}_{i},\boldsymbol{0};\sigma_{p};\mathrm{free}\,\rangle|^{2},\label{eq:5.4}
\end{equation}
or
\begin{equation}
\frac{d\sigma}{d\Omega}|_{\mathrm{C}}=\frac{p^{2}}{16\sigma_{p}^{4}}|\langle\,f\,|\,i\,\rangle|^{2}.\label{eq:5.5}
\end{equation}

There is a simple way to see how the powers of $\sigma_{p}$ and $p$
arise. The area of the incoming wavepacket perpendicular to the average
momentum direction scales as $A\sim\sigma_{x}^{2}\sim1/\sigma_{p}^{2}.$
The final state solid angle element scales as $\Delta\Omega\sim(\sigma_{p}/p)^{2}.$
Then the probability of the event is
\begin{eqnarray}
P & \sim & \frac{1}{A}\frac{d\sigma}{d\Omega}\Delta\Omega,\label{eq:5.8}
\end{eqnarray}
so
\begin{eqnarray}
\frac{d\sigma}{d\Omega} & \sim & \frac{p^{2}}{\sigma_{p}^{4}}P.\label{eq:5.10}
\end{eqnarray}

\section{Evaluation of the finite amplitude and extraction of the scattering
amplitude}

We now have the finite amplitude, using (\ref{eq:4.32},\ref{eq:4.33})
and $d_{00}^{l}(\theta)=P_{l}(\cos\theta),$ a Legendre polynomial,
\begin{equation}
\langle\,f\,|\,i\,\rangle=\int_{0}^{\infty}dk\sum_{l=0}^{\infty}|\Psi_{\mathrm{free}}(k,l)|^{2}\,e^{i\varphi(k,l)}\,P_{l}(\cos\theta)\{1+\mathcal{O}(\sqrt{\epsilon})\},\label{eq:7.2}
\end{equation}
where the phases are
\begin{equation}
\varphi(k,l)=2kR-\frac{k^{2}T}{2m_{0}}+2\sigma_{l}(k).\label{eq:7.4}
\end{equation}

The time shift phase factor is
\begin{equation}
e^{-ik^{2}T/2m_{0}}=e^{-ip^{2}T/2m_{0}}e^{-i\beta T(k-p)}(1+\mathcal{O}(\sqrt{\epsilon})),\label{eq:7.5}
\end{equation}
with $\beta=p/m_{0}$ the average speed of the wavepacket.

We also expand the Coulomb phase shifts in powers of $k-p$:
\begin{equation}
\sigma_{l}(k)=\sigma_{l}(p)+(k-p)\frac{\partial\sigma_{l}(p)}{\partial k}+\mathcal{R}(k,l).\label{eq:7.5.1}
\end{equation}
A numerical investigation shows that the remainder, $\mathcal{R}(k,l),$
is negligible on our parameter space.

So we can find the leading approximation to the $k$ integral for
small $\epsilon$,
\begin{equation}
\int_{0}^{\infty}dk\,e^{-(k-p)^{2}/2\sigma_{p}^{2}}e^{-i(k-p)\Delta_{l}(t)}\sim\sqrt{2\pi\sigma_{p}^{2}}\,e^{-\Delta_{l}^{2}(t)/8\sigma_{x}^{2}}\label{eq:7.5.5}
\end{equation}
(with negligible error from extending the lower limit of the integral
to $-\infty$), to give
\begin{equation}
|\langle\,f\,|\,i\,\rangle|=|\sum_{l=0}^{\infty}\frac{(l+\frac{1}{2})e^{-2\sigma_{p}^{2}(l+\frac{1}{2})^{2}/p^{2}}}{(p/2\sigma_{p})^{2}}e^{-\Delta_{l}^{2}(T)/8\sigma_{x}^{2}}\,e^{i2\sigma_{l}(p)}P_{l}(\cos\theta)|\{1+\mathcal{O}(\sqrt{\epsilon})\},\label{eq:7.6}
\end{equation}
with
\begin{equation}
\Delta_{l}(T)=\beta T-2R-2\frac{\partial\sigma_{l}(p)}{\partial k}.\label{eq:7.6.5}
\end{equation}

Taking into account the spatial shifts caused by the logarithmic phase,
the time it would take a free particle to start at radius $R-\eta(p)(\ln(2pR)-1)/p,$
be deflected at the origin without time shift and finish at radius
$R-\eta(p)(\ln(2pR)-1)/p$ would be $T_{\mathrm{free}}$ with
\begin{equation}
\beta T_{\mathrm{free}}=2R-2\frac{\eta(p)}{p}(\ln(2pR)-1).\label{eq:7.6.6}
\end{equation}
For the actual interacting evolution, we define the spatial displacement
($\beta$ times the time delay) as a function of the shift time, $T$,
in units of $\sigma_{x},$ as
\begin{align}
\delta(T) & \equiv\frac{\beta T-\beta T_{\mathrm{free}}}{\sigma_{x}}\label{eq:7.6.7}\\
 & =\frac{\beta T-2R+2\eta(p)(\ln(2pR)-1)/p}{\sigma_{x}}.\label{eq:7.6.8}
\end{align}
Then the Gaussian factor dependent on $T$ becomes
\begin{equation}
e^{-\Delta_{l}^{2}(T)/8\sigma_{x}^{2}}=e^{-(\delta(T)-\xi_{l}(p))^{2}/8}\label{eq:7.6.9}
\end{equation}
with
\begin{equation}
\xi_{l}(p)\equiv4\epsilon\eta(p)\{\ln(2pR)-1-\frac{\partial\sigma_{l}[\eta(p)]}{\partial\eta}\}.\label{eq:7.7.3-1}
\end{equation}
We used
\begin{equation}
\frac{\partial\sigma_{l}(p)}{\partial k}=-\frac{\eta(p)}{p}\frac{\partial\sigma_{l}[\eta(p)]}{\partial\eta}.\label{eq:7.7.5}
\end{equation}
The form on the right-hand side is more advantageous for computation.

Note that if the interaction made any change to the free spreading
formula (\ref{eq:3.6}), it would be caused by a phase proportional
to $(k-p)^{2}.$ Since in the expansions of the logarithmic phase
(\ref{eq:4.23}) and the Coulomb phase shift (\ref{eq:7.5.1}), we
found those second-order terms to be negligible with our choice of
parameters, there is effectively no spreading over the course of the
experiment.

Now we numerically evaluate the series in
\begin{equation}
P(\theta,\delta)=|\langle\,f\,|\,i\,\rangle|^{2}=4\epsilon^{4}|\sum_{l=0}^{\infty}(2l+1)e^{-2\epsilon^{2}(l+\frac{1}{2})^{2}}e^{-(\delta-\xi_{l}(p))^{2}/8}e^{i2\sigma_{l}(p)}P_{l}(\cos\theta)|^{2}\label{eq:7.8}
\end{equation}
to find the probability of the process. In practice we scan in $\delta$
to find the peak. The equation for the differential cross section
is then
\begin{equation}
\frac{d\sigma}{d\Omega}(\theta,\delta)=\frac{1}{4p^{2}}|\sum_{l=0}^{\infty}(2l+1)e^{-2\epsilon^{2}(l+\frac{1}{2})^{2}}e^{-(\delta-\xi_{l}(p))^{2}/8}e^{i2\sigma_{l}(p)}P_{l}(\cos\theta)|^{2},\label{eq:7.8.5}
\end{equation}
with the warning, discussed earlier, that if this result produces
a peak of width $\Delta\theta\sim\epsilon$ around the forward direction,
we must use the methods of Section VIII to separate the peak from
the scattering.

Note that this formula (\ref{eq:7.8}) can be used for any short-range
central potential with the replacements
\begin{align}
\sigma_{l}(p) & \rightarrow\delta_{l}(p),\label{eq:7.9}\\
\xi_{l}(p) & \rightarrow\frac{2}{\sigma_{x}}\frac{\partial\delta_{l}(p)}{\partial k},\label{eq:7.10}
\end{align}
where $\delta_{l}(k)$ are the phase shifts for that potential and
there are no logarithmic phase contributions to $\xi_{l}(p).$

\section{Results and discussion of measurability}

Now we calculate the probability $P(\theta,\delta)$ as given by (\ref{eq:7.8},\ref{eq:7.7.3-1},\ref{eq:4.2-1}),
for several cases.

We choose the momentum resolution parameter $\epsilon=0.001$ mostly
for illustration purposes, as it is much less than unity, and to limit
computation time. There is a sense in which the ``true'' differential
cross section is the limit of our results as $\epsilon\rightarrow0^{+}.$
However, we will find that there is not uniform convergence in scattering
angle, $\theta,$ or strength parameter, $\eta.$ Yet $\sigma_{p}$
is a physical parameter, the width in momentum of the wavepacket.
To provide a realistic description of a scattering experiment would
require an estimate of this parameter.

We consider radioactive and accelerator sources. If alpha particles
were emitted from Radium-226 with a momentum spread determined by
the natural linewidth (the reciprocal of the decay lifetime as determined
by the Heisenberg uncertainty principle) the value of $\epsilon$
would be so small (of order $10^{-33}$) that deviations from the
Rutherford formula could never be observed. However physical effects
broaden the linewidth from a solid sample. Again for radioactive emission
of alpha particles, using extremely thin samples to reduce broadening
effects, linewidths as small as $\Delta E=2\,\mathrm{keV}$ have been
observed \cite{Pomme2015}. For the $E=4.8\,\mathrm{MeV}$ emission
of Radium-226, this would give $\epsilon<2.1\times10^{-4},$ assuming
that the actual energy width of the wavepackets could be smaller than
the observed linewidth.

In accelerator experiments, the energy resolution of the beam is well
controlled, so it should be possible to put an upper bound on $\epsilon.$
This matter requires further consideration for a specific experiment.

In Figure 2 we plot the angular dependence of the probability for
the particular choice $\eta=10,$ then the dependence of the probability
on $\eta$ for the scattering angle $\theta=\pi/2.$ In both cases
we choose $\delta=0,$ not attempting to find the maxima of the probabilities.
For comparison we plot the Rutherford ``probability'',
\begin{equation}
P_{\mathrm{Ruth}}(\theta,\eta,\epsilon)=\frac{4\epsilon^{4}\eta^{2}}{\sin^{4}\frac{\theta}{2}},\label{eq:73}
\end{equation}
which we obtain by naïvely converting the Rutherford differential
cross section (\ref{eq:I.6}) into a probability using (\ref{eq:5.5}).
Of course this is not a true probability since it rises above unity
and diverges at $\theta=0.$

In Figure 2(a) we see excellent agreement with the Rutherford formula
over almost the entire range of angles, but a deviation at low angles.
The probability we calculate cannot rise above unity, so a deviation
is to be expected. In Figure 2(b) we see deviations from the Rutherford
formula at high energies (low $\eta$) and at low energies (large
$\eta$). We assume that the first of these is due to numerical error,
but we do not investigate further in this paper. The deviation at
low energies will be investigated shortly with a special treatment
of the phase shifts, where we conclude that there is a physical effect.

\begin{figure}
\begin{centering}
\includegraphics[width=12cm]{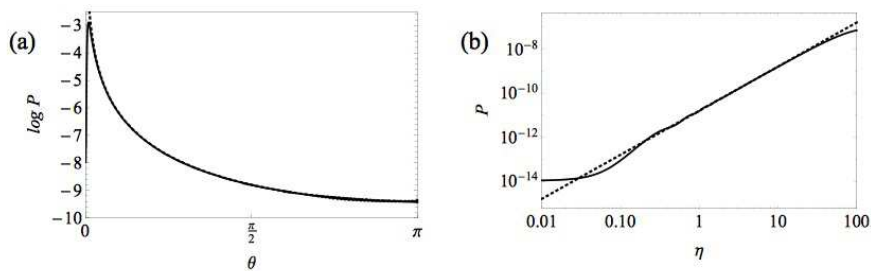}
\par\end{centering}
\caption{Probabilities (solid) for (a) $\eta=10$ as a function of $\theta$
and (b) $\theta=\pi/2$ as a function of $\eta.$ In both cases the
Rutherford probability (dotted) is shown for comparison.}

\end{figure}

We find that the timing varies significantly with scattering angle.
First we calculate the probability in the forward direction, $\theta=0,$
for $\eta=10,$ as a function of the scaled spatial shift $\delta$.
This shows a maximum at $\delta=+1.2,$ corresponding to a time delay.
We find later that the peak probability as a function of angle occurs
at around $\theta=0.03.$ The profile in $\delta$ for this angle
is shown in Figure 3(a). We find the maximum probability for this
angle is at $\delta=+0.4,$ corresponding to a time delay. We will
discuss the measurability of time shifts shortly. Then we calculate
the angular distribution, $P(\theta,+0.4),$ for this value of $\delta$,
shown in Figure~3(b). We see deviations from the Rutherford formula
for scattering angles less than about $0.1\,\mathrm{rad}=5.7\textdegree.$
There is a shadow zone of low probability of angular width $\sim0.01\,\mathrm{rad}=0.6\textdegree$
around the forward direction.

\begin{figure}
\begin{centering}
\includegraphics[width=12cm]{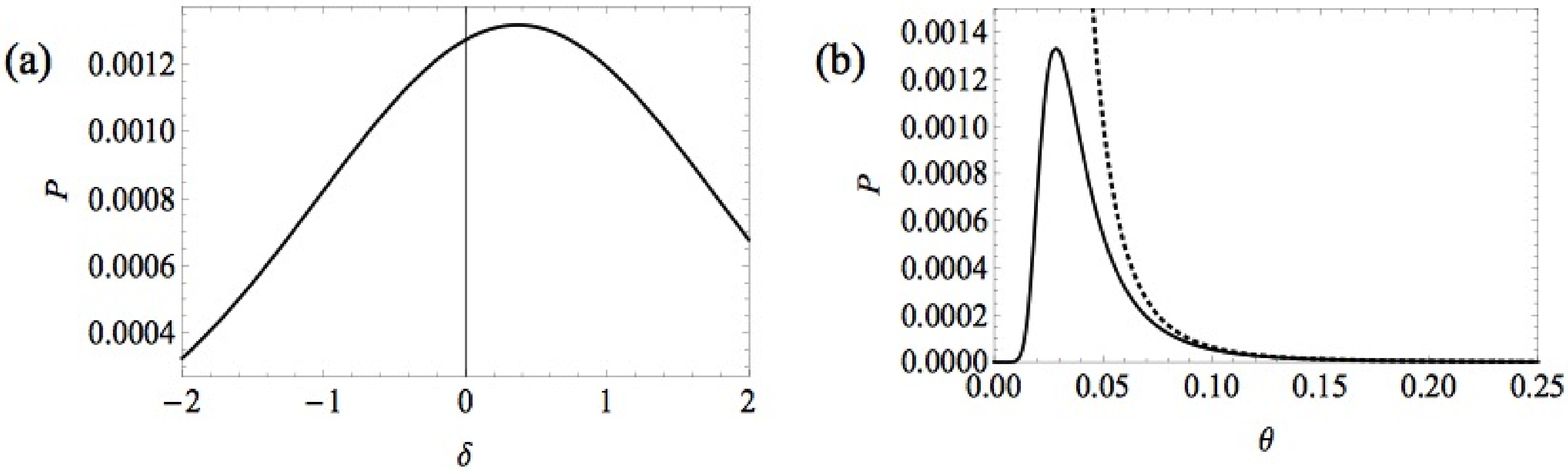}
\par\end{centering}
\caption{Probabilities (solid) for $\eta=10:$ (a) $P(0.03,\delta)$ and (b)
$P(\theta,+0.4)$ (solid) compared with $P_{\mathrm{Ruth}}(\theta,\eta,\epsilon)$
(dotted).}

\end{figure}

We have found that as $\epsilon$ is decreased, the region of violation
of the Rutherford formula becomes smaller (in approximately direct
proportion).

We repeat this procedure for an attractive interaction ($\eta=-10$)
and find the profile in $\delta$ for scattering angle $\theta=0.03$
shown in Figure 4(a). We see a peak at $\delta=-0.4,$ corresponding
to an advancement in time. Then we set $\delta=-0.4$ and obtain the
probability shown in Figure 4(b). The results are very similar to
the case $\eta=+10.$ The Rutherford probability for these two cases
($\eta=\pm10$) is identical.

\begin{figure}
\begin{centering}
\includegraphics[width=12cm]{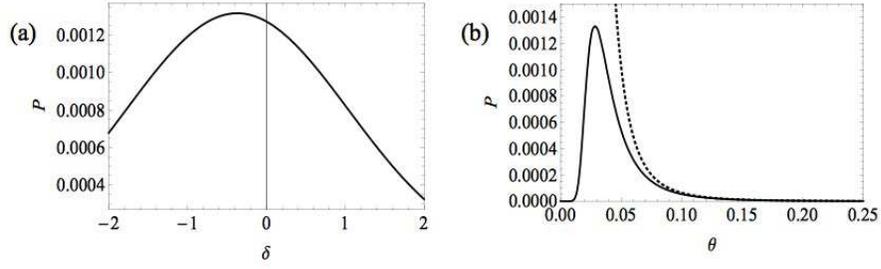}
\par\end{centering}
\caption{Probabilities (solid) for $\eta=-10:$ (a) $P(0.03,\delta)$ and (b)
$P(\theta,-0.4)$ compared with $P_{\mathrm{Ruth}}(\theta,\eta,\epsilon)$
(dotted).}
\end{figure}

We consider weaker interactions ($\eta=1,0.1$) with probability results
shown in Figure 5, only focussing on low angles. For $\eta=1$ we
see a strong forward peak and $P(0,\delta)$ takes its maximum at
$\delta=0.05.$ For $\eta=0.1$ we find $P(0,\delta)$ takes its maximum
at $\delta=0.0$. The probability profile in Figure 5(b) is essentially
that expected for no interaction, proportional to $\exp(-\theta^{2}/4\epsilon^{2})$
(dot-dashed) with a peak probability very close to unity. Thus we
see more deviations from the Rutherford formula, at low interaction
strengths. The calculation shown in Figure 2(b) for low $\eta$ is
attempting to calculate an extremely small probability at $\theta=\pi/2$,
leading us to the conclusion that we are encountering numerical errors
in this region.

\begin{figure}
\begin{centering}
\includegraphics[width=12cm]{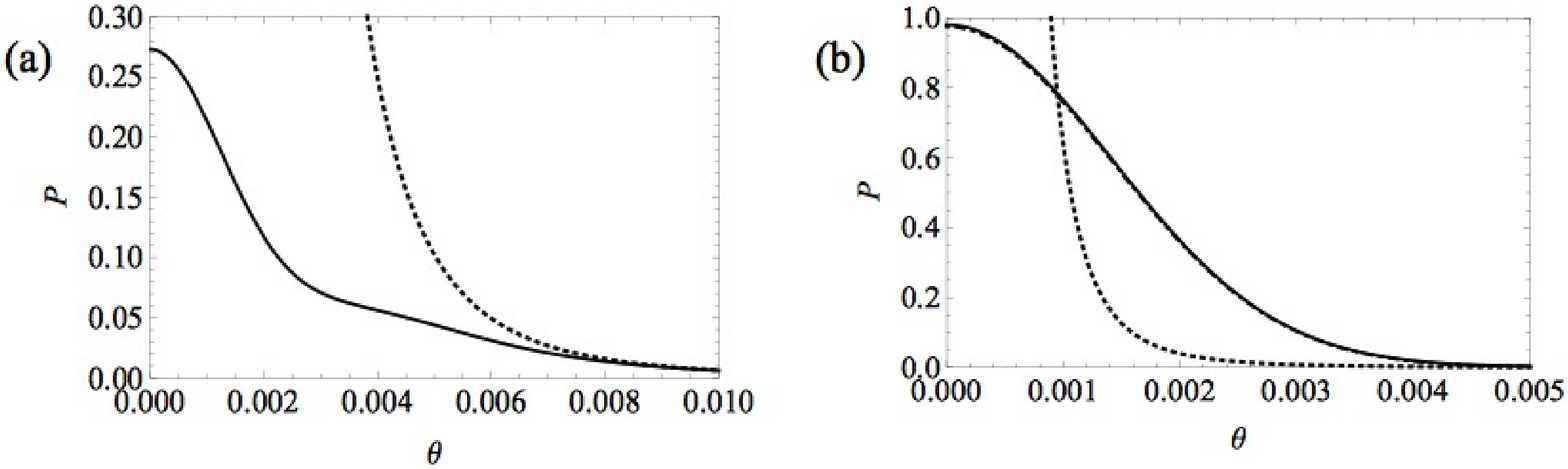}
\par\end{centering}
\caption{(a) Probability $P(\theta,0.05)$ for $\eta=1$ and (b) probability
$P(\theta,0)$ for $\eta=0.1$ compared to $P(0,0)\exp(-\theta^{2}/4\epsilon^{2})$
(dot-dashed). Both are compared to $P_{\mathrm{Ruth}}(\theta,\eta,\epsilon)$
(dotted).}

\end{figure}

Our numerical investigations and the result (\ref{eq:8.7}) that we
obtain in Section VIII suggest that the angular width of the shadow
zone will increase as $\eta$ increases (incident momentum, $p,$
decreases). To investigate scattering at very large values of $\eta,$
for numerical stability in our computation, it was necessary to obtain
an asymptotic approximation of the Coulomb phase shifts,
\begin{equation}
e^{i2\sigma_{l}[\eta]}=e^{i2\eta\{\ln(\sqrt{(l+1)^{2}+\eta^{2}})-1\}}e^{i(2l+1)\phi_{l}(\eta)}\{1+\mathcal{O}(\frac{1}{|l+1+i\eta|})\},\label{eq:74}
\end{equation}
with $\phi_{l}(\eta)=\tan^{-1}(\eta/(l+1)).$ This then leads to
\begin{equation}
\xi_{l}[\eta]=4\epsilon\eta\{\ln(2pR)-1-\ln(\sqrt{(l+1)^{2}+\eta^{2}})-1\}.\label{eq:75}
\end{equation}

The experiment we envision, that we wish to describe with this calculation,
is the scattering of alpha particles ($Z_{2}=+2$) by a thin gold
($Z_{1}=+79$) foil, similar to the original experiments of Rutherford,
Geiger and Marsden \cite{Rutherford1911,Geiger1909}. The energies
of alpha particles emitted by radioactive sources are too large for
our purposes. The $4.8\,\mathrm{MeV}$ emission of Radium-226 with
$\epsilon=2.1\times10^{-4}$ gives $\eta=23$ and $\theta_{0}=1.1\textdegree,$
a very small deviation region that would be obscured by the unscattered
particles. So we suggest that a helium nucleus accelerator \cite{Anokhin2013}
could be configured to produce sufficiently low energies.

In Section IV we imposed a bound on the strength parameter, $|\eta|\leq844,$
to satisfy physical constraints. So we consider first the case $\eta=800$
($E=3.8\,\mathrm{keV}$ with the above experimental parameters). Figure
6 shows the predicted differential cross section compared to the Rutherford
formula (\ref{eq:1.6}). We see a cross section dominated by backscattering,
with a shadow zone almost to $\theta=\pi/2$, in agreement with the
Rutherford formula only at $\theta=\pi.$ Even with a beam width of
$10\textdegree=0.17\,\mathrm{rad}$ obscuring the forward direction
(see discussion at the end of this Section), this shadow zone would
be clearly observable. We find $P(\pi/2,\delta)$ takes its maximum
at $\delta=+5.3.$

\begin{figure}
\begin{centering}
\includegraphics[width=8.6cm]{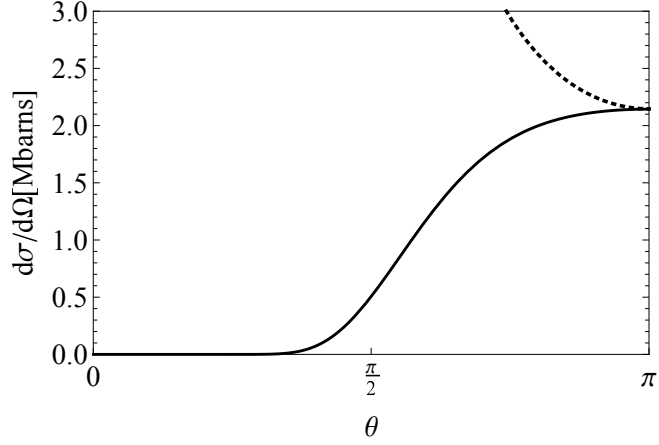}
\par\end{centering}
\caption{Predicted differential cross section (solid) compared to the Rutherford
formula (dotted) for the parameters given in the text and incident
energy $E=3.8\,\mathrm{keV}$.}
\end{figure}

We examine the range of energies over which this shadow zone could
be observed. We choose an observation angle of $\theta=\pi/4.$ There
should no difficulty ensuring a beam width sufficiently narrow that
the essentially unscattered particles do not obscure the signal. For
this scattering angle, we calculate the ratio of predicted to Rutherford
differential cross sections
\begin{equation}
\rho(E)=\frac{d\sigma}{d\Omega}(\frac{\pi}{4},\delta_{\mathrm{max}}(E);E)/\frac{Z_{1}^{2}Z_{2}^{2}\alpha^{2}}{16E^{2}\sin^{4}\frac{\pi}{8}}.\label{eq:6}
\end{equation}
We sampled the scaled spatial shifts at several energies and found
a best fit $\delta_{\mathrm{max}}(E).$ The results are shown in Figure
7. We see that the ratio rises rapidly with energy. The potentially
most definitive measurements would be on the region $3.8\leq E[\mathrm{keV}]\leq20.$
We note $\rho(3.8\,\mathrm{keV})=2.5\times10^{-7}.$

\begin{figure}[H]
\begin{centering}
\includegraphics[width=8.6cm]{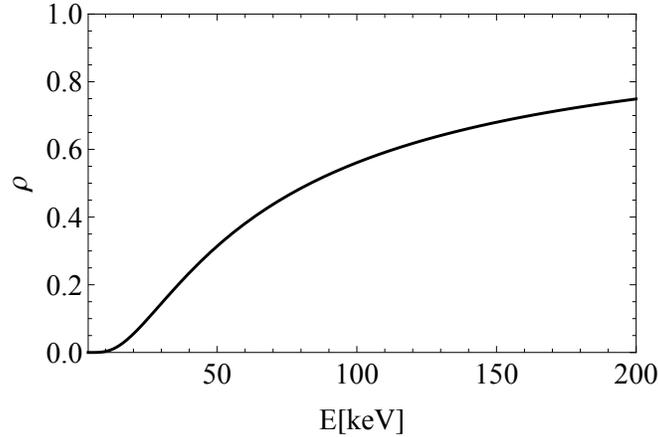}
\par\end{centering}
\caption{Ratio of differential cross sections for $3.8\leq E[\mathrm{keV}]\leq200.$}
\end{figure}

Since this example has the largest scaled spatial shift encountered,
$\delta=+5.3,$ we calculate the time delay in seconds for this case
and find $\Delta t=2.3\times10^{-16}\,\mathrm{s},$ not easily measured.

Wigner \cite{Wigner1955} reasoned that the spatial shift for $l$-wave
scattering would be twice the derivative of the phase shift with respect
to $k$. This is precisely what we have found, after taking into account
the time shifts caused by the logarithmic phase. Wigner was considering
in what way the classical concept of causality could be extended to
the quantum domain, but his reasoning was based on the assumption
of a potential of finite range.

The form of our results should be cause for concern. In addition to
the violations of the accepted Rutherford formula at low angles, we
see (except at low energies) no strong forward peak as usually expected
in a scattering experiment and the very low scattering in the forward
direction raises questions about violation of the optical theorem.

On the first of these points, our scattering model has a special geometry,
that of a head-on collision of wavepackets. In a real scattering experiment,
the projectiles approach the targets with a distribution of impact
parameters, displacements perpendicular to the incoming momentum direction.
The physical expectation is that wavepackets with sufficiently large
impact parameters will pass through essentially unscattered. These
events will build up the strong, narrow peak around zero scattering
angle that we expect. We will examine the effects of nonzero impact
parameters on Coulomb scattering in a future work.

We consider the optical theorem as applied to our system in the next
Section.

\section{Conservation of probability and the optical theorem}

We find a simple integral relation satisfied by the probabilities
$P(\theta,\delta)$ that expresses conservation of total probability.
Starting from (\ref{eq:7.8}) and using the orthogonality of the Legendre
polynomials and the normalization integral for a shifted Gaussian,
it is easily shown that
\begin{equation}
\frac{1}{2\epsilon^{2}}\frac{1}{\sqrt{4\pi}}\int_{-\infty}^{\infty}d\delta\int_{0}^{\pi}\sin\theta\,d\theta\,P(\theta,\delta)=8\epsilon^{2}\sum_{l=0}^{\infty}(l+\frac{1}{2})\,e^{-4\epsilon^{2}(l+\frac{1}{2})^{2}}=1+\mathcal{O}(\epsilon^{2}).\label{eq:8.3}
\end{equation}
This relation holds for all interactions, including the free case
with
\begin{equation}
P_{\mathrm{Free}}(\theta,\delta)=e^{-\theta^{2}/4\epsilon^{2}}e^{-\delta^{2}/4}.\label{eq:8.4}
\end{equation}

For the Coulomb interaction, we have cases with no narrow peak with
probability close to unity around the forward direction that we would
expect for short-range potentials. So the relation (\ref{eq:5.5})
between the probability and the differential cross section can be
used for all angles. Then we find another integral relation
\begin{equation}
\frac{1}{\sqrt{4\pi}}\int_{-\infty}^{\infty}d\delta\int_{0}^{\pi}2\pi\sin\theta\,d\theta\,\frac{d\sigma}{d\Omega}=\pi\sigma_{x}^{2}.\label{eq:8.4.1}
\end{equation}
We hesitate to call this the ``total cross section'' for the Coulomb
interaction, as it is independent of the strength parameter, $\eta,$
and diverges as the momentum width $\sigma_{p}\rightarrow0^{+}.$
For all potentials we find the sum rule
\begin{equation}
\frac{1}{\sqrt{4\pi}}\int_{-\infty}^{\infty}d\delta\int_{0}^{\pi}2\pi\sin\theta\,d\theta\,F(\theta,\delta)=\pi\sigma_{x}^{2},\label{eq:8.4.2}
\end{equation}
with $F(\theta,\delta)=p^{2}P(\theta,\delta)/16\sigma_{p}^{4}$ including
the forward peak and the scattering. The forward peak dominates this
normalization integral for short-range potentials, which does not
give the total scattering cross section. A separation of the forward
peak and the scattering is necessary to define the differential scattering
cross section in these cases, as we do below.

Our result for the Coulomb case $\eta=800$ differed most dramatically
from the expected form for short-range potentials. We would like to
be confident that the total probability in our final states is unity.
We found a best fit for $\delta_{\mathrm{max}}(\theta)$ after finding
the values of $\delta$ where the probability took its maximum value
for several sampled values of $\theta.$ Then we approximated the
integral over $\theta$ as a discrete sum with 200 intervals and found
\begin{equation}
\frac{1}{2\epsilon^{2}}\int_{0}^{\pi}\sin\theta\,d\theta\,P(\theta,\delta_{\mathrm{max}}(\theta))\cong0.998.\label{eq:8.4.5}
\end{equation}

We can use the integral relation (\ref{eq:8.3}) to find an estimate
of the angular size of the region where the Rutherford formula is
violated. In our numerical investigations, we have found that the
probability factorizes in the form
\begin{equation}
P(\theta,\delta)=e^{-(\delta-\delta_{\mathrm{max}}(\theta))^{2}/4}P_{\mathrm{max}}(\theta),\label{eq:8.5}
\end{equation}
to a very good approximation. Clearly the way to extract the maximum
probability is by the integral
\begin{equation}
\frac{1}{\sqrt{4\pi}}\int_{-\infty}^{\infty}d\delta\,P(\theta,\delta)=P_{\mathrm{max}}(\theta).\label{eq:8.5.1}
\end{equation}
Then suppose the shadow region had vanishing probability, $P_{\mathrm{max}}(\theta),$
on $0\leq\theta\leq\theta_{0}$ and took the Rutherford form (\ref{eq:73})
at higher angles. Then we would have the integral relation
\begin{equation}
\frac{1}{2\epsilon^{2}}\int_{\theta_{0}}^{\pi}\sin\theta\,d\theta\,\frac{4\epsilon^{4}\eta^{2}}{\sin^{4}\frac{\theta}{2}}=1.\label{eq:8.6}
\end{equation}
For $\epsilon\eta\ll1,$ which includes the cases other than $\eta=800$
that we have been considering, we find
\begin{equation}
\theta_{0}(\epsilon,\eta)\cong4\epsilon|\eta|.\label{eq:8.7}
\end{equation}
This gives $\theta_{0}(0.001,\pm10)\cong0.04$, $\theta_{0}(0.001,1)\cong0.004$.
From Figures 3 and 4, these estimates are seen to give good characterizations
of the regions of violation. For the case $\eta=0.1,$ there is a
narrow forward peak instead of a shadow zone, so an estimate of $\theta_{0}$
is not appropriate.

We derive an optical theorem using partial wave analysis (as done
by Newton \cite{Newton1982}), modified to take into account amplitudes
that depend on $\delta$ as well as $\theta.$ We start by defining
an amplitude with a definite phase, as necessary for a relation between
an amplitude and a phase-independent total cross section, by
\begin{equation}
A(\theta,\delta)=2\epsilon^{2}\sum_{l=0}^{\infty}(2l+1)e^{-2\epsilon^{2}(l+\frac{1}{2})^{2}}e^{-(\delta-\xi_{l}(p))^{2}/8}e^{i2\sigma_{l}(p)}P_{l}(\cos\theta),\label{eq:8.8}
\end{equation}
so that $P(\theta,\delta)=|A(\theta,\delta)|^{2}.$ Then we apply
the identity
\begin{equation}
e^{i2\sigma_{l}(p)}=1+i2e^{i\sigma_{l}(p)}\sin\sigma_{l}(p),\label{eq:8.9}
\end{equation}
with the intention of separating the amplitude into a part that describes
the narrow forward peak and a remaining part that we identify as the
scattering. This is equivalent to separating the transition matrix,
$T,$ from the $S$ matrix by $S=1+iT.$ This gives
\begin{equation}
A(\theta,\delta)=A_{\mathrm{F}}(\theta,\delta)+iA_{\mathrm{S}}(\theta,\delta),\label{eq:8.10}
\end{equation}
with
\begin{align}
A_{\mathrm{F}}(\theta,\delta) & =2\epsilon^{2}\sum_{l=0}^{\infty}(2l+1)e^{-2\epsilon^{2}(l+\frac{1}{2})^{2}}e^{-(\delta-\xi_{l}(p))^{2}/8}P_{l}(\cos\theta),\label{eq:8.11}\\
A_{\mathrm{S}}(\theta,\delta) & =4\epsilon^{2}\sum_{l=0}^{\infty}(2l+1)e^{-2\epsilon^{2}(l+\frac{1}{2})^{2}}e^{-(\delta-\xi_{l}(p))^{2}/8}e^{i\sigma_{l}(p)}\sin\sigma_{l}(p)P_{l}(\cos\theta).\label{eq:8.12}
\end{align}

Using this separation, we define a scattering amplitude, a function
of $\theta$ only, by integrating over $\delta$:
\begin{equation}
f(\theta)=\frac{p}{4\sigma_{p}^{2}}\frac{1}{\sqrt{8\pi}}\int_{-\infty}^{\infty}d\delta\,A_{\mathrm{S}}(\theta,\delta),\label{eq:8.13}
\end{equation}
with the prefactor chosen to be consistent with (\ref{eq:5.5}). Then
we find
\begin{equation}
\mathrm{Im}\,f(0)=\frac{1}{p}\sum_{l=0}^{\infty}(2l+1)e^{-2\epsilon^{2}(l+\frac{1}{2})^{2}}\sin^{2}\sigma_{l}(p).\label{eq:8.14}
\end{equation}
Similarly, we define the total cross section as another integral over
$\delta$ and an integral over all scattering directions:
\begin{align}
\sigma & =\frac{1}{\sqrt{4\pi}}\int_{-\infty}^{\infty}d\delta\int_{0}^{\pi}2\pi\sin\theta\,d\theta\,\frac{p^{2}}{16\sigma_{p}^{4}}|A_{\mathrm{S}}(\theta,\delta)|^{2}\label{eq:8.15}\\
 & =\frac{4\pi}{p^{2}}\sum_{l=0}^{\infty}(2l+1)e^{-4\epsilon^{2}(l+\frac{1}{2})^{2}}\sin^{2}\sigma_{l}(p).\label{eq:8.16}
\end{align}

Now we point out the great distinction between the Coulomb case and
that of a short-range potential. In the latter case, $A_{\mathrm{F}}(\theta,\delta)$
converges to a narrow function peaked at $\theta=0$ with height less
than unity by $\mathcal{O}(\epsilon^{2}).$ (These results were tested
numerically on the finite-range spherical square well potential \cite{Messiah1961}.)
The sums in (\ref{eq:8.14}) and (\ref{eq:8.16}) converge without
the need of the Gaussians in $l+1/2,$ to
\begin{align}
\sigma_{\mathrm{S-R}} & =\frac{4\pi}{p^{2}}\sum_{l=0}^{\infty}(2l+1)\sin^{2}\sigma_{l}(p)\{1+\mathcal{O}(\epsilon^{2})\},\label{eq:8.16.1}\\
\mathrm{Im}\,f_{\mathrm{S-R}}(0) & =\frac{1}{p}\sum_{l=0}^{\infty}(2l+1)\sin^{2}\sigma_{l}(p)\{1+\mathcal{O}(\epsilon^{2})\}.\label{eq:8.16.2}
\end{align}
The scattering amplitude, $f_{\mathrm{S-R}}(\theta)$ is finite for
all angles (for the spherical square well) and has no narrow peak
at $\theta=0,$ so the separation (\ref{eq:8.9}) achieves the intended
goal of separating the forward peak from the scattering. We note that
there is a computational benefit in using the form (\ref{eq:8.12})
instead of (\ref{eq:8.8}) if we are not interested in calculating
the forward peak: the sums over $l$ can be taken to a cutoff value
much lower than when the form (\ref{eq:8.8}) is used.

Then the optical theorem,
\begin{equation}
\sigma_{\mathrm{S-R}}=\frac{4\pi}{p}\mathrm{Im}\,f_{\mathrm{S-R}}(0),\label{eq:8.20}
\end{equation}
follows immediately from (\ref{eq:8.16.1}) and (\ref{eq:8.16.2}).

We see that our method can be recast into the familiar results of
partial wave analysis, for short-range potentials, with two benefits:
the regularization of the forward peak and the ability to calculate
time shifts.

For the Coulomb potential, taking $\eta=10$ as an example, we find
that \textit{both} $A_{\mathrm{F}}(\theta,\delta)$ and the imaginary
part of $A_{\mathrm{S}}(\theta,\delta)$ have narrow peaks around
$\theta=0,$ both of the form
\begin{equation}
A_{\mathrm{F}}(\theta,\delta)\cong\mathrm{Im}\,A_{\mathrm{S}}(\theta,\delta)\sim e^{-\theta^{2}/8\epsilon^{2}}e^{-(\delta-\delta_{0})^{2}/8}\quad\mathrm{for}\ |\theta|\apprle10\epsilon,\label{eq:8.21}
\end{equation}
while the real part of $A_{\mathrm{S}}(\theta,\delta)$ takes much
smaller values. So using these separated amplitudes, the observed
low value of the probability close to $\theta=0$ comes about partly
from a near cancellation of two very similar terms in
\begin{equation}
P(\theta,\delta)=(A_{\mathrm{F}}(\theta,\delta)-\mathrm{Im}\,A_{\mathrm{S}}(\theta,\delta))^{2}+(\mathrm{Re}\,A_{\mathrm{S}}(\theta,\delta))^{2}.\label{eq:8.22}
\end{equation}
We see that the separation (\ref{eq:8.9}), in the $\eta=10$ Coulomb
case, does not separate the amplitude into a forward peak and a scattering
term, justifying our choice not to apply it.

Even the case $\eta=0.1,$ which shows a forward peak in the probability
close to unity, gives unintuitive results when the separation (\ref{eq:8.9})
is applied. In that case all three of $A_{\mathrm{F}}(\theta,\delta),$
$\mathrm{Im}\,A_{\mathrm{S}}(\theta,\delta)$ and $\mathrm{Re}\,A_{\mathrm{S}}(\theta,\delta)$
show narrow peaks close to unity.

As for the possibility of an optical theorem in the Coulomb case,
we plot the ratio
\[
\gamma(\eta)=\frac{\sigma_{\mathrm{C}}}{4\pi\mathrm{Im}\,f_{\mathrm{C}}(0)/p}=\frac{\sum_{l=0}^{\infty}(2l+1)e^{-4\epsilon^{2}(l+\frac{1}{2})^{2}}\sin^{2}\sigma_{l}(p)}{\sum_{l=0}^{\infty}(2l+1)e^{-2\epsilon^{2}(l+\frac{1}{2})^{2}}\sin^{2}\sigma_{l}(p)}
\]
for a range of values of $\eta$ (and for $\epsilon=0.001$) in Figure
8. We see that nowhere does it take the value unity. Furthermore,
we have seen that the quantities in the ratio do not have the physical
meanings of their counterparts for short-range potentials.

\begin{figure}
\begin{centering}
\includegraphics[width=8cm]{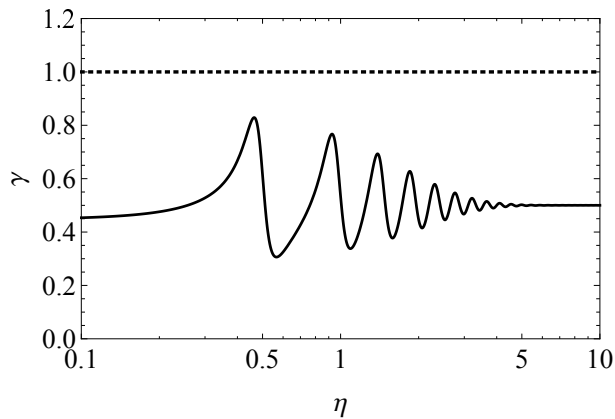}
\par\end{centering}
\caption{Optical theorem ratio $\gamma(\eta)$ (solid) for $\epsilon=0.001,$
compared with unity (dotted).}

\end{figure}

\section{Comparison with other works}

Boris \textit{et al.} \cite{Boris1993} numerically evolved wavepackets
in a Coulomb potential. For an initial wavepacket state vector, $|\,\psi(0)\,\rangle$,
they numerically calculated the amplitudes on the interacting basis
vectors $|\,k,l,m\,\rangle$,
\begin{equation}
\langle\,k,l,m\,|\,\psi(0)\,\rangle=\int d^{3}r\,\langle\,k,l,m\,|\,\boldsymbol{r}\,\rangle\langle\,\boldsymbol{r}\,|\,\psi(0)\,\rangle,\label{eq:I.5}
\end{equation}
where the $\langle\,\boldsymbol{r}\,|\,k,l,m\,\rangle$ are simply
related to the energy and angular momentum eigenvectors of the full
Coulomb Hamiltonian (by (\ref{eq:2.2.5})). Then they plotted the
position probability density, $|\psi(\boldsymbol{r},t)|^{2},$ where
\begin{equation}
\psi(\boldsymbol{r},t)=\int_{0}^{\infty}dk\sum_{l=0}^{\infty}\sum_{m=-l}^{l}\langle\,\boldsymbol{r}\,|\,k,l,m\,\rangle e^{-ik^{2}t/2m_{0}}\langle\,k,l,m\,|\,\psi(0)\,\rangle.\label{eq:I.6}
\end{equation}
However, they did not extract the differential cross section or calculate
time shifts.

Their Figure 15, for the scattering of a wavepacket by a repulsive
Coulomb potential, shows a shadow zone around the forward direction.
However that was for a wavepacket with a nonzero impact parameter,
as they were investigating whether wavepackets would follow classical
trajectories. Since they did not consider the head-on (zero impact
parameter) case, direct comparison with our results is not possible.
Also, their wavepackets spread noticeably over the course of their
numerical experiments, so they appear not to be in the regime $\epsilon\ll1.$
It would be of value to repeat their simulations with our parameters
to allow direct comparison.

We are also interested in whether wavepackets follow classical trajectories,
and will investigate this in a future work.

Several authors \cite{Nutt1968,Dettmann1971} have used wavepackets
to deal with the divergence of the $T$ matrix in formal scattering
theory applied to the Coulomb interaction \cite{Schwinger1964}. Dettmann
\cite{Dettmann1971} was able to reproduce the Rutherford formula
but, crucially, he did not calculate the modulus-squared of the $T$
matrix elements for final momenta close to the initial momentum. It
would be of value to repeat his calculation in that region to allow
direct comparison with our results.

\section{Conclusions}

The inclusion of wavepackets into the description of a scattering
experiment does not merely smear the scattering amplitude over a small
angular width. Instead, for Coulomb scattering, we predict a finite
probability of scattering into the forward direction, where the Rutherford
formula diverges. Generally we find excellent agreement with the Rutherford
formula except for small angles around the forward direction. We conclude
that these deviations would not have been observable in previous experiments.
However, consideration of low incident energies shows the deviation
to be magnified and raises the possibility of experimental observation
of a shadow zone.

We note that the existence of a shadow zone is consistent with the
view of a scattering process as the diffraction of a wave, in the
appropriate parameter regime.

A wavepacket treatment also implies, and allows the calculation of,
time shifts in the detection probabilities. However, these are generally
very small for Coulomb scattering.

From a theoretical perspective, a wavepacket description allows the
technique of partial wave analysis from scattering theory to be applied
to the long-range Coulomb potential. We use a formalism that allows
treatment of the logarithmic phase in (\ref{eq:4.1-1}), a consequence
of the infinite range of the Coulomb potential. This phase alters
the long-distance trajectory of the wavepacket, in close agreement
with the corresponding classical trajectory.

Our method involves the calculation of \textit{probabilities} for
wavepacket to wavepacket transitions and then relates these to the
differential cross section. That fact that a probability must be finite
leads directly to our prediction of deviations from the Rutherford
formula for the differential cross section.

This method is made possible by the availability of the exact solutions
for the energy angular momentum eigenfunctions. To apply the method
to other central potentials would require the phase shifts for those
potentials. Phase shifts have been calculated for the Yukawa \cite{Hamzavi2012},
Morse \cite{Rawitscher2002} and spherical well \cite{Messiah1961}
potentials, for example. Note that Rawitscher \textit{et al. }\cite{Rawitscher2002}
calculated the motion of a wavepacket at a resonance using a partial
wave method. There are numerical methods available to calculate phase
shifts in other cases \cite{Klozenberg1974}.

We note that partial wave analysis is widely used in nuclear physics,
where a Coulomb interaction is often present in addition to short-range
nuclear forces \cite{Bertulani2004}. The Coulomb amplitude (\ref{eq:1.1})
is usually inserted to avoid the divergent partial wave series. The
method presented here may allow the inclusion of the Coulomb force
into the partial wave analysis.

The method could be extended to two-particle scattering by forming
the eigenvectors of total energy, momentum and centre-of-mass frame
angular momentum \cite{Jacob1959,Macfarlane1962}. Particles with
spin are not an obstacle. If relativistic interactions are to be investigated,
it is not widely known that there are relativistic probability amplitudes
for particles of any spin, as discussed by Fong and Rowe \cite{Fong1968}.
For the electron, these are two-component rather than four-component
objects.

\bibliographystyle{vancouver}

\begin{acknowledgments}
The author is grateful to Diane and Lydia and to his mother, Mark,
Adam and their families for their love and support. The author is
also grateful to Anwen King for helpful discussions, and to Yao-Zhong
Zhang and Ian Marquette for their support.

Two referees pointed out matters that needed attention, for which
the author is also grateful.

The author receives financial support from a UQ Research Scholarship.
\end{acknowledgments}

\end{document}